\documentclass[10pt]{ismd08}
\usepackage{graphicx}
\usepackage{cite,./mcite}

\usepackage{cite}
\usepackage{amsmath}
\usepackage{graphicx}
\usepackage{epsfig}
\usepackage{hyperref}
\usepackage{url}
\usepackage{amssymb}

\newcommand{\lp}{\left(}
\newcommand{\rp}{\right)}

\newcommand{\JD}{\mathrm{JD}}

\begin{document}
\title{Review on recent developements in jet finding}
\author{Juan Rojo}
\institute{LPTHE, UPMC -- Paris 6 and
    Paris-Diderot -- Paris 7, CNRS UMR 7589,
    Paris (France) \\ INFN, Sezione di Milano, Via Celoria 16,
I - 20133, Milano (Italy)}
\maketitle
\begin{abstract}
We review recent developements related to jet clustering
algorithms and jet finding. 
These include fast implementations
of sequential recombination algorithms,
 new IRC safe algorithms, quantitative
determination of jet areas and
quality measures for jet finding, among many others.
We also briefly discuss the status of jet finding in heavy ion collisions,
where full QCD jets have been measured 
for the first time at RHIC.
\end{abstract}

\paragraph{Recent developements in jet algorithms}
With the upcoming start-up of the LHC,  
jet finding techniques have received considerable attention.
In this brief review, we outline some of the most important
developements in jet algorithms and related subjects in the
recent years.
Much more detailed reviews
can be found in \cite{Buttar:2008jx,Ellis:2007ib}.

An important developement  has been  the
fast implementation of the $k_T$ \cite{Catani:1993hr}
and Cambridge/Aachen \cite{Catani:1991hj,Wobisch:1998wt} 
jet algorithms.
Prior to 2005, existing
implementations scaled as $N^3$, with $N$ the number
of particles to be clustered, thus making it unpractical
for high multiplicity collisions like $pp$ at the LHC and even more
in Heavy Ions Collisions (HIC).
Thanks to computational geometry methods, the performance
of these algorithms was made to scale as $N\ln N$ \cite{Cacciari:2005hq}. 
These fast implementations are available through 
the {\tt FastJet} package \cite{fastjet}, together
with area-based subtraction methods and plugins to external
jet finders (see below).

\begin{figure}[ht!]
\begin{center}
\includegraphics[width=0.73\textwidth]{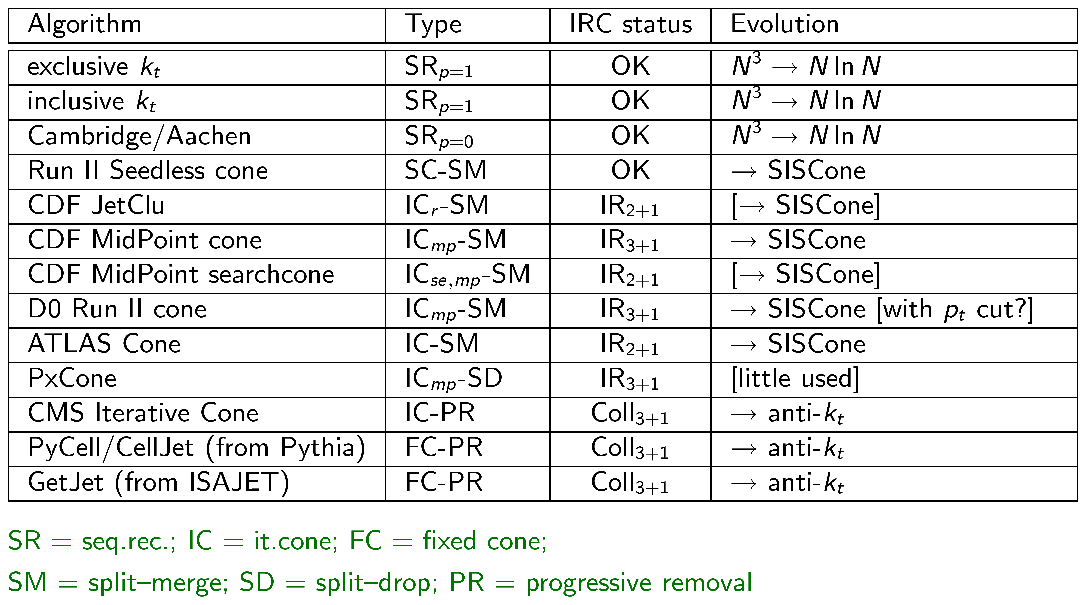}
\caption{\small Summary of the progress in jet algorithms since
2005.}
\label{fig:jetlist}
\end{center}
\end{figure}

Another important achievement has been
the formulation of a practical (scaling as $N^2 \ln N$)
 infrared and collinear (IRC) safe
 cone algorithm, SISCone \cite{Salam:2007xv}. 
Unlike all other commonly used
cone algorithms, SISCone is IRC safe
to all orders in perturbation theory by construction. 
This property allows one to compare any perturbative computation
with experimental data, which for IRC unsafe algorithms
is impossible beyond some fixed order, indicated
in Fig. \ref{fig:jetlist}.
As discussed in \cite{Salam:2007xv}, the phenomenological
implications of SISCone when compared with the (IRC unsafe)
 commonly used MidPoint cone algorithm range from  few percent
differences in the inclusive jet spectrum, somewhat larger
in the presence of realistic Underlying Event (UE), up to 50\% differences
for more exclusive observables, like the tails of jet-mass spectra
in multi-jet events.

\begin{figure}[ht!]
\begin{center}
\includegraphics[width=0.38\textwidth]{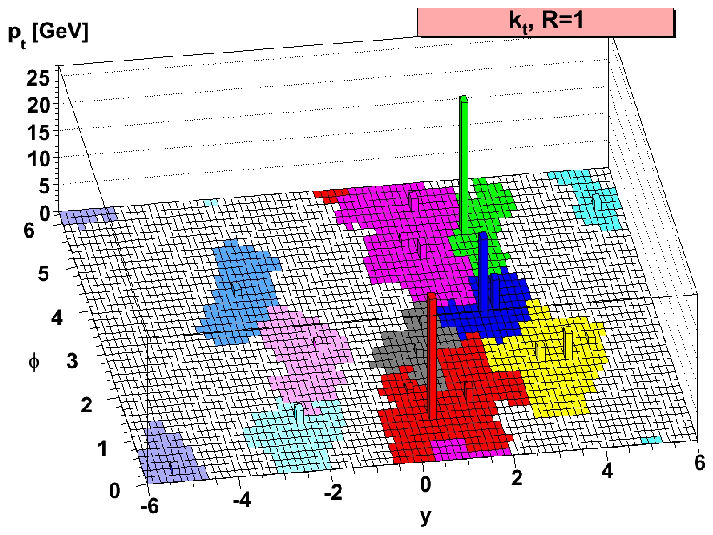}
\includegraphics[width=0.38\textwidth]{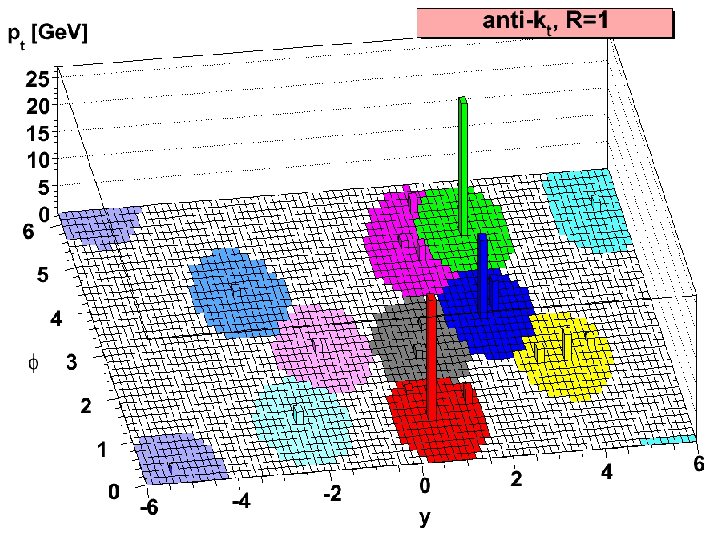}
\caption{\small Jet areas for the $k_t$ (left) and anti-$k_t$ 
(right) algorithms for
$R=1$.}
\label{fig:plot-areas}
\end{center}
\end{figure}

There has been historically some confusion  about the
concept of the {\it size} of a jet, specially since the
naive jet area
is ambiguous beyond LO. The situation  was recently clarified by the
introduction of quantitative definitions of jet areas
based on the {\it catchment} properties of hard jets with respect 
to very soft
particles, called {\it ghosts} in
\cite{Cacciari:2008gn}.  Examples of jet areas
defined with such a technique are shown in Fig. 
\ref{fig:plot-areas}. On top of their
theoretical importance, jet areas have important
applications related to the subtraction of
soft backgrounds coming from the UE or from Pile-Up (PU),
 both in $pp$ and in
$AA$ collisions, as discussed in \cite{Cacciari:2007fd}.

Another recently developed IRC safe
jet algorithm is the anti-$k_t$ 
algorithm \cite{Cacciari:2008gp}. This algorithm
is related to  $k_T$ and Cam/Aa  by its
distance measure,  $  d_{ij} \equiv { \min(k_{ti}^{ 2p 
},k_{tj}^{2p })} 
{ \Delta R_{ij}^2/R^2}$,
with $p=-1$ ($p=1$ corresponds to $k_T$ and $p=0$ to Cam/Aa).
The anti-$k_T$ algorithm  has the property of
being soft-resilient, that is,
due to its distance soft particles are always clustered
with hard particles first.
This property leads to rather regular jet areas,
which become perfectly circular in the limit in which all
hard particles are separated in the $(y,\phi)$ plane
by at least a distance $R$, as
can be seen in Fig. \ref{fig:plot-areas}.
Another important advantage of the anti-$k_t$  algorithm is
that it has a very small back-reaction \cite{Cacciari:2008gn},
that is, the presence of a soft background has reduced effects
on which hard particles are clustered into a given jet.

The recent progress in jet algorithms 
can be summarized in Fig. \ref{fig:jetlist}.
Each IRC unsafe cone jet algorithm can now
be replaced by the corresponding IRC safe one, with a similar
physics performance, shown in the last
column of Fig. \ref{fig:jetlist}.
SISCone is the natural IRC safe replacement for 
MidPoint-type iterative 
cone algorithms with split-merge (IC-SM), while anti-$k_T$
is so 
for iterative cone algorithms of the progressive removal (IC-PR) type
\cite{Buttar:2008jx}.

This brief review is unable to cover
many other  interesting developements related
to jets and jet finding in the recent years. 
Some of those not discussed here
 include the use of jet substructure as a useful technique to
improve signal significance in various channels
at the LHC (see for example 
\cite{Butterworth:2008iy,Kaplan:2008ie,Thaler:2008ju}), 
analytical studies of the interplay
between perturbative and non-perturbative effects
in jet finding \cite{Dasgupta:2007wa}, the infrared safe
definition of jet flavour and its application 
to precision predictions for $b-$jets at hadron
 colliders \cite{Banfi:2006hf,Banfi:2007gu} or the
impact of jet measurements, both at the Tevatron and at HERA,
 in global analysis of PDFs \cite{Ball:2008by,Nadolsky:2008zw}.

\paragraph{Performance of jet algorithms at LHC}
A recurring question in jet studies is ``what is the best jet
definition for a given specific analysis''? 
Most existing techniques either 
use as a reference  unphysical Monte Carlo partons (an ambiguous
concept beyond LO)
and/or assume some shape for the measured kinematical distributions.
To overcome these disadvantages, 
a new strategy to quantify the performance of jet definitions in
kinematic reconstruction tasks has been recently
introduced \cite{jet-performance},  which is designed to make use exclusively
of physical observables.

In Ref. \cite{jet-performance} two quality measures respecting
the above requirements are proposed, and applied to the kinematic
reconstruction of invariant mass distributions in dijet events
for a wide range of energies.
These quality measures can in turn be mapped into an effective luminosity
ratio, defined as
\begin{equation}
  \label{eq:rhol_basic_def}
  \rho_{\cal L}(\JD_2 / \JD_1) \equiv 
  \frac{{\cal L}(\text{needed with }\JD_2)}
       {{\cal L}(\text{needed with }\JD_1)} 
  = \left[
    \frac{\Sigma\lp \JD_1 \rp} {\Sigma\lp \JD_2 \rp} \right]^2 \, .
\end{equation}
Given a certain signal significance $\Sigma$ with 
jet definition $\JD_2$, $\rho_{\cal
  L}(\JD_2/\JD_1)$ indicates the factor more luminosity needed to
obtain the same significance as with jet definition $\JD_1$.

\begin{figure}
\begin{center}
\includegraphics[width=0.625\textwidth]{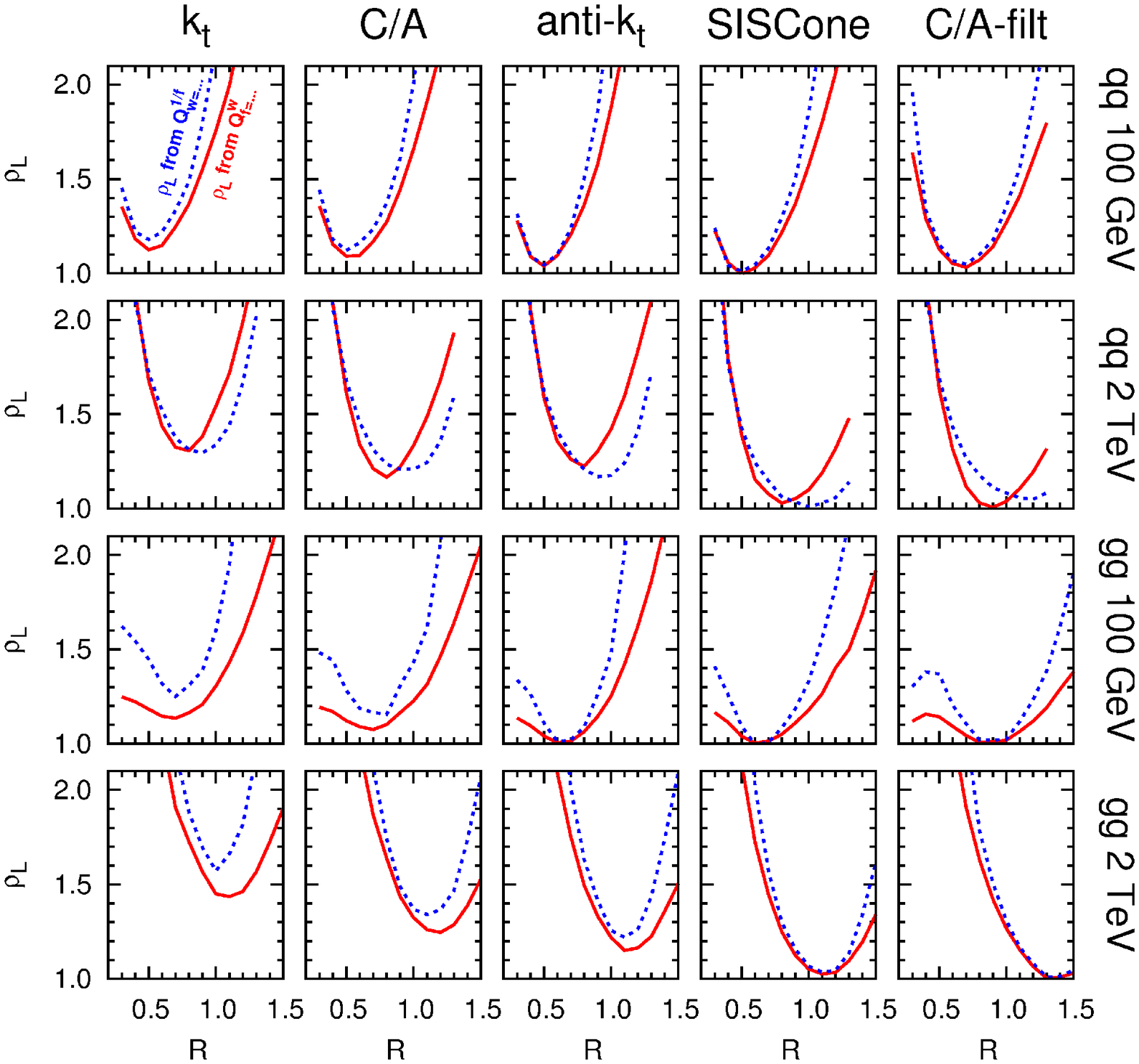}
\caption{}{\small 
The effective luminosity ratio, Eq.
 \ref{eq:rhol_basic_def}, for quark and gluon jets
at 100 GeV and 2 TeV 
\cite{jet-performance}.
\label{fig:rhoL-summary} }
\end{center}
\end{figure}

The results of \cite{jet-performance}
 over a large range
of jet definitions,\footnote{There results can also be accessed through an
interactive web tool \cite{jet-performance-web} which allows
the user to compare the jet finding quality for a wide range of 
parameters (jet algorithm, $R$, value of PU, ...).} 
summarized in Fig. \ref{fig:rhoL-summary},
indicate that for gluon jets, and in general for TeV scales, there
are significant benefits to be had from using larger radii that
those commonly used, up to
$R\gtrsim 1$. In general, SISCone and C/A-filt (Cam/Aa supplemented
with a filtering procedure \cite{Butterworth:2008iy}) show the
best performance. These conclusions are robust in the presence
of high-luminosity PU, when subtracted with the
jet area technique \cite{Cacciari:2007fd}.

\paragraph{Jet finding in AA collisions at RHIC and LHC}
While QCD jets are ubiquitous in pp collisions,
until this year no real jet reconstruction had been obtained
 in the much more
challenging environment of HIC. Indeed, usually in HIC one refers to
the leading particle of the event as a {\it jet}. However, 
reconstructing full QCD
jets provides a much more precise window to the
properties of the hot and dense
medium created in the collision than just leading
particles.

\begin{figure}
\begin{center}
\includegraphics[width=0.56\textwidth]{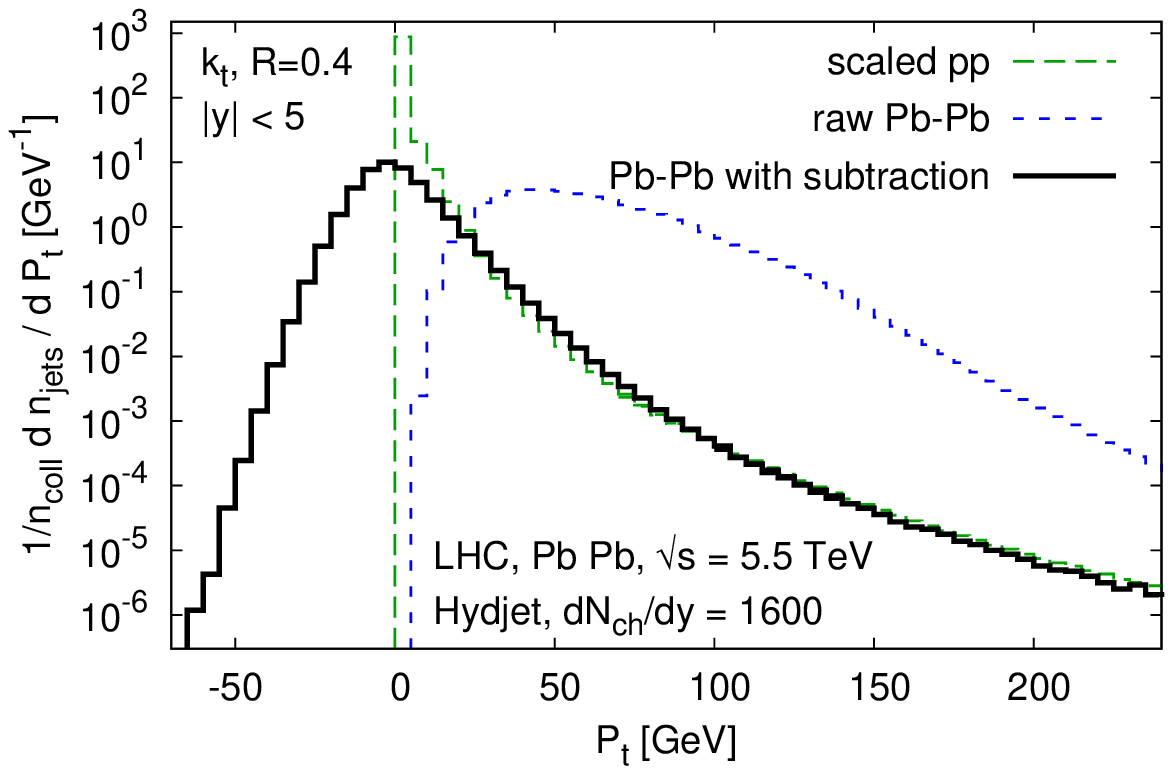}
\includegraphics[width=0.43\textwidth]{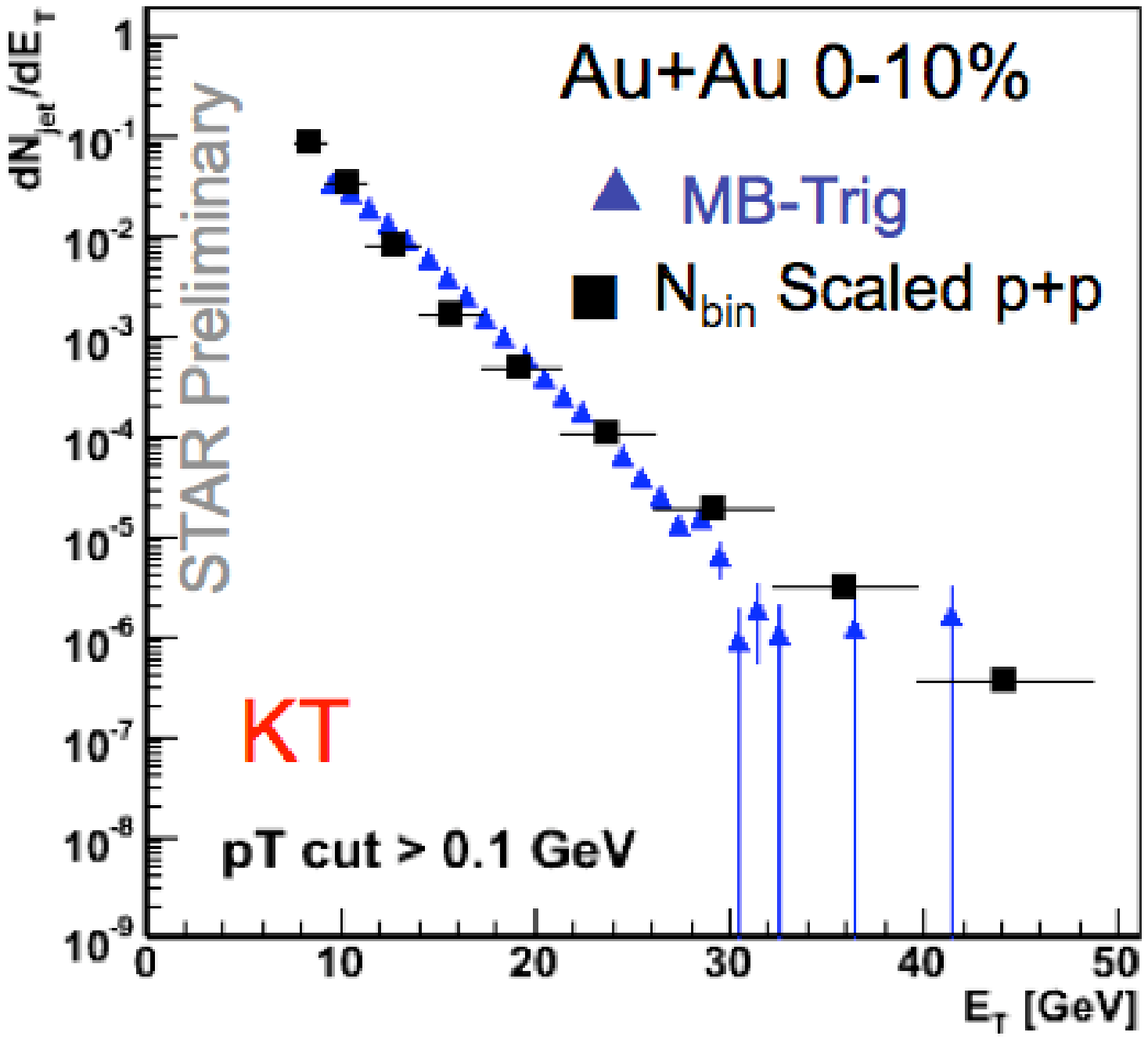}
\caption{}{\small Left: the simulated inclusive jet spectrum at the
LHC with the $k_T$ algorithm, including subtraction, from
\cite{Cacciari:2007fd}. Right: the inclusive jet spectrum
measured with $k_T$ by STAR at RHIC, from \cite{Salur:2008hs}.}
\label{fig:plot-hic}
\end{center}
\end{figure}

The difficulty in
 reconstructing jets in HIC stems from the huge backgrounds,
which need to be subtracted in order to compare with baseline results.
There are various techniques to subtract such large backgrounds. 
In \cite{Cacciari:2007fd} it was shown how the
{\tt FastJet} jet area method could efficiently subtract 
such backgrounds in HIC at the LHC with a good accuracy 
(see Fig. \ref{fig:plot-hic}).

A major breakthrough in jet finding was  the  recent first
 measurement of QCD jets in HIC
by the STAR collaboration at RHIC \cite{Salur:2008hs}.
In Fig. \ref{fig:plot-hic} we show their measurement with the
$k_T$ algorithm.
These results
should have important consequences for understanding 
the medium properties in HIC.

It would be important, after these initial measurements,
to improve the control on the accuracy of the
subtraction procedure, as well
as to understand the differences between the performances
of different jet
algorithms. Ongoing studies \cite{hic}
suggest that one of the important
sources of systematic error in the HIC jet reconstruction
is back-reaction \cite{Cacciari:2008gn}, therefore anti-$k_t$ is potentially
interesting in this situation due to its small back-reaction
\cite{Cacciari:2008gp}. 
Ref. \cite{hic} also investigates how the use of local ranges for the
determination of the background level $\rho$ might help reducing the
effects of point-to-point background fluctuations.
However, more work is still required in order to determine the optimal
settings for jet finding in HIC.

\paragraph{Outlook}

Jet finding has seen
a large number of important developements in the recent years,
However, there is still
room for more progress, which should be driven by the actual
requirements of LHC data analysis. Jet finding will also be essential
to exploit the heavy-ion program at the LHC 
as proved by the latest RHIC jet measurements.

\paragraph{Acknowledgments}
This work has been supported by the grant ANR-05-JCJC-0046-01 (France).
The author wants to acknowledge M. Cacciari, G. Salam and
G. Soyez for help and material
while preparing this review.

\begin{footnotesize}
\providecommand{\etal}{et al.\xspace}
\providecommand{\href}[2]{#2}
\providecommand{\coll}{Coll.}
\catcode`\@=11
\def\@bibitem#1{%
\ifmc@bstsupport
  \mc@iftail{#1}%
    {;\newline\ignorespaces}%
    {\ifmc@first\else.\fi\orig@bibitem{#1}}
  \mc@firstfalse
\else
  \mc@iftail{#1}%
    {\ignorespaces}%
    {\orig@bibitem{#1}}%
\fi}%
\catcode`\@=12

\bibliographystyle{ismd08} 

\begin{mcbibliography}{10}

\bibitem{Buttar:2008jx}
C.~Buttar {\em et al.}~(2008).
\newblock \href{http://www.arXiv.org/abs/0803.0678 [hep-ph]}{{\tt 0803.0678
  [hep-ph]}}\relax
\relax
\bibitem{Ellis:2007ib}
S.~D. Ellis, J.~Huston, K.~Hatakeyama, P.~Loch, and M.~Tonnesmann,
\newblock Prog. Part. Nucl. Phys.{} {\bf 60},~484~(2008).
\newblock \href{http://www.arXiv.org/abs/0712.2447}{{\tt 0712.2447}}\relax
\relax
\bibitem{Catani:1993hr}
S.~Catani, Y.~L. Dokshitzer, M.~H. Seymour, and B.~R. Webber,
\newblock Nucl. Phys.{} {\bf B406},~187~(1993)\relax
\relax
\bibitem{Catani:1991hj}
S.~Catani, Y.~L. Dokshitzer, M.~Olsson, G.~Turnock, and B.~R. Webber,
\newblock Phys. Lett.{} {\bf B269},~432~(1991)\relax
\relax
\bibitem{Wobisch:1998wt}
M.~Wobisch and T.~Wengler~(1998).
\newblock \href{http://www.arXiv.org/abs/hep-ph/9907280}{{\tt
  hep-ph/9907280}}\relax
\relax
\bibitem{Cacciari:2005hq}
M.~Cacciari and G.~P. Salam,
\newblock Phys. Lett.{} {\bf B641},~57~(2006).
\newblock \href{http://www.arXiv.org/abs/hep-ph/0512210}{{\tt
  hep-ph/0512210}}\relax
\relax
\bibitem{fastjet}
M.~Cacciari, G.~P. Salam, and G.~Soyez~(2005-2008).
\newblock
  \href{http://www.lpthe.jussieu.fr/~salam/fastjet}{
{\tt http://www.lpthe.jussieu.fr/$\sim$salam/fastjet}}\relax
\relax
\bibitem{Salam:2007xv}
G.~P. Salam and G.~Soyez,
\newblock JHEP{} {\bf 05},~086~(2007).
\newblock \href{http://www.arXiv.org/abs/0704.0292 [hep-ph]}{{\tt 0704.0292
  [hep-ph]}}\relax
\relax
\bibitem{Cacciari:2008gn}
M.~Cacciari, G.~P. Salam, and G.~Soyez,
\newblock JHEP{} {\bf 04},~005~(2008).
\newblock \href{http://www.arXiv.org/abs/0802.1188}{{\tt 0802.1188}}\relax
\relax
\bibitem{Cacciari:2007fd}
M.~Cacciari and G.~P. Salam,
\newblock Phys. Lett.{} {\bf B659},~119~(2008).
\newblock \href{http://www.arXiv.org/abs/0707.1378}{{\tt 0707.1378}}\relax
\relax
\bibitem{Cacciari:2008gp}
M.~Cacciari, G.~P. Salam, and G.~Soyez,
\newblock JHEP{} {\bf 04},~063~(2008).
\newblock \href{http://www.arXiv.org/abs/0802.1189}{{\tt 0802.1189}}\relax
\relax
\bibitem{Butterworth:2008iy}
J.~M. Butterworth, A.~R. Davison, M.~Rubin, and G.~P. Salam,
\newblock Phys. Rev. Lett.{} {\bf 100},~242001~(2008).
\newblock \href{http://www.arXiv.org/abs/0802.2470}{{\tt 0802.2470}}\relax
\relax
\bibitem{Kaplan:2008ie}
D.~E. Kaplan, K.~Rehermann, M.~D. Schwartz, and B.~Tweedie,
\newblock Phys. Rev. Lett.{} {\bf 101},~142001~(2008).
\newblock \href{http://www.arXiv.org/abs/0806.0848}{{\tt 0806.0848}}\relax
\relax
\bibitem{Thaler:2008ju}
J.~Thaler and L.-T. Wang,
\newblock JHEP{} {\bf 07},~092~(2008).
\newblock \href{http://www.arXiv.org/abs/0806.0023}{{\tt 0806.0023}}\relax
\relax
\bibitem{Dasgupta:2007wa}
M.~Dasgupta, L.~Magnea, and G.~P. Salam,
\newblock JHEP{} {\bf 02},~055~(2008).
\newblock \href{http://www.arXiv.org/abs/0712.3014}{{\tt 0712.3014}}\relax
\relax
\bibitem{Banfi:2006hf}
A.~Banfi, G.~P. Salam, and G.~Zanderighi,
\newblock Eur. Phys. J.{} {\bf C47},~113~(2006).
\newblock \href{http://www.arXiv.org/abs/hep-ph/0601139}{{\tt
  hep-ph/0601139}}\relax
\relax
\bibitem{Banfi:2007gu}
A.~Banfi, G.~P. Salam, and G.~Zanderighi,
\newblock JHEP{} {\bf 07},~026~(2007).
\newblock \href{http://www.arXiv.org/abs/0704.2999}{{\tt 0704.2999}}\relax
\relax
\bibitem{Ball:2008by}
{ NNPDF} Collaboration, R.~D. Ball {\em et al.}~(2008).
\newblock \href{http://www.arXiv.org/abs/0808.1231}{{\tt 0808.1231}}\relax
\relax
\bibitem{Nadolsky:2008zw}
P.~M. Nadolsky {\em et al.},
\newblock Phys. Rev.{} {\bf D78},~013004~(2008).
\newblock \href{http://www.arXiv.org/abs/0802.0007}{{\tt 0802.0007}}\relax
\relax
\bibitem{jet-performance}
M.~Cacciari, J.~Rojo, G.~P. Salam, and G.~Soyez~(2008).
\newblock \href{http://www.arXiv.org/abs/0810.1304}{{\tt 0810.1304}}\relax
\relax
\bibitem{jet-performance-web}
M.~Cacciari, J.~Rojo, G.~P. Salam, and G.~Soyez~(2008),
\newblock \href{http://quality.fastjet.fr/}{{\tt
  http://quality.fastjet.fr/}}\relax
\relax
\bibitem{Salur:2008hs}
{ STAR} Collaboration, S.~Salur~(2008).
\newblock \href{http://www.arXiv.org/abs/0809.1609}{{\tt 0809.1609}}\relax
\relax
\bibitem{hic}
M.~Cacciari, J.~Rojo, G.~P. Salam, and G.~Soyez~, in preparation\relax
\relax
\end{mcbibliography}
\end{footnotesize}
\end{document}